\documentclass[journal]{IEEEtran}

\usepackage[labelsep=period]{caption}
\usepackage{makecell}
\usepackage{float}
\usepackage{caption}
\usepackage{subcaption}
\usepackage{longtable}
\usepackage{tabularx}
\usepackage{tabu}
\usepackage{multirow}
\usepackage{tikz}
\usepackage{graphicx}
\usepackage{tikzscale}
\usepackage{pgfplots}
\pgfplotsset{compat=1.17}
\usepackage{setspace}
\usepackage{booktabs}
\usepackage{siunitx}
\usepackage{adjustbox}
\usepackage{soul, color}
\usepackage{xcolor}

\usepackage[most]{tcolorbox}
\newtcolorbox{highlighted}{colback=yellow,coltext=red,breakable}

\usepackage{algorithmic}
\usepackage{amsmath,amssymb,amsfonts}
\usepackage[ruled,linesnumbered]{algorithm2e}
\usepackage[font=small,labelfont=bf]{caption}
\SetKwFor{Foreach}{for each}{do}{endForEach}%
\SetKwFor{For}{for}{do}{endfor for}%
\SetKwIF{uIf}{ElseIf}{Else}{if}{then}{else if}{else}{endIf if}%
\SetKwFor{While}{while}{do}{endWhile while}%

\begin{document}

\title{
Smart Policy Control for Securing Federated Learning Management System
}

\author{%
Aditya Pribadi Kalapaaking, Ibrahim Khalil, and Mohammed Atiquzzaman
}

\maketitle

\begin{abstract}
The widespread adoption of Internet of Things (IoT) devices in smart cities, intelligent healthcare systems, and various real-world applications have resulted in the generation of vast amounts of data, often analyzed using different Machine Learning (ML) models. Federated learning (FL) has been acknowledged as a privacy-preserving machine learning technology, where multiple parties cooperatively train ML models without exchanging raw data. However, the current FL architecture does not allow for an audit of the training process due to the various data-protection policies implemented by each FL participant. Furthermore, there is no global model verifiability available in the current architecture. This paper proposes a smart contract-based policy control for securing the Federated Learning (FL) management system. First, we develop and deploy a smart contract-based local training policy control on the FL participants' side. This policy control is used to verify the training process, ensuring that the evaluation process follows the same rules for all FL participants. We then enforce a smart contract-based aggregation policy to manage the global model aggregation process. Upon completion, the aggregated model and policy are stored on blockchain-based storage. Subsequently, we distribute the aggregated global model and the smart contract to all FL participants. Our proposed method uses smart policy control to manage access and verify the integrity of machine learning models. We conducted multiple experiments with various machine learning architectures and datasets to evaluate our proposed framework, such as MNIST and CIFAR-10.
\end{abstract}


%
\IEEEpeerreviewmaketitle

\section{Introduction}
The Internet of Things (IoT) has been involved in various services, including smart cities, smart healthcare, and smart manufacturer, to enhance the quality of life, the efficiency of urban services, operation, and competitiveness \cite{9903398}. A distributed network made up of IoT devices connected via wired or wireless networks continuously interacts with the outside world to provide a variety of data sources, including images, text, video, and other sorts of data. The distributed network also enables efficient and effective sharing of IoT data resources and information. However, due to the substantial amount of data generated by IoT sensors, an intelligent system is necessary to operate the system autonomously, as IoT devices are typically resource-limited and cannot independently execute machine learning algorithms. With the aid of edge computing, IoT clusters can form intelligent networks when combined with machine learning \cite{9766403}. However, having an effective machine learning model necessitates extensive data from many IoT clusters, which is often difficult to collect and utilize due to privacy concerns, security risks, and other associated challenges \cite{9997105}.

A new type of distributed machine learning approach called federated learning (FL) controls the training process without storing the data in the server \cite{bonawitz2019towards}. FL is a privacy-preserving distributed machine learning protocol that reduces data communication costs by employing a model-first strategy. This strategy allows centralized servers to maintain a global model and transmit its parameters to connected devices and systems instead of gathering large datasets. The edge devices then perform the local training process using the global model received from the server and send the trained local model back for a global model aggregation process. As a result,  edge computing can have a good machine learning model without ever sending data to the cloud or to an external party \cite{9352033}.

In recent studies, \cite{bonawitz2019towards} proposed federated learning to enable multiple participants to collaboratively train a model by exchanging local model updates with a parameter server. This method is more secure than centralized training as machine learning models learn from IoT data without relying on a third-party cloud to keep their data \cite{yu2021research}. While federated learning guarantees the privacy of local data by generating a global model without sharing data among participants, current FL architecture still faces numerous challenges, such as model integrity and transparency \cite{9403374}. Furthermore, an adversary could tamper with the local or global model and result in misclassification.

Several existing defense methods mainly focus on securing the training process using encrypted training \cite{zhang2020batchcrypt} or leveraging a trusted execution environment (TEE) \cite{9155414} to perform the training process. However, the current approach is inefficient since the training process requires significant computation resources and takes a long time. Moreover, the aggregation server and participants cannot validate the machine learning model they receive. Hence, an auditable and verifiable machine learning management system is needed to enhance the security of federated learning architecture.

Blockchain is a distributed system that links data structure for data storage, ensuring the data is resistant to modification and tampering \cite{monrat2019survey}. Initially, blockchain applications were mainly limited to cryptocurrencies and financial transactions. However, with the development of smart contracts, blockchain technology has opened up a range of new applications \cite{zou2019smart}. Smart contracts are self-executing contracts that are triggered when certain conditions are met, enforcing the rules of the agreement between parties. Once deployed to a blockchain network, smart contracts are immutable and tamperproof, providing a secure, transparent, and efficient way to conduct business in a decentralized, trustless environment \cite{9737334}.

This paper proposes a secure federated learning management system utilizing smart policy control. We introduce a smart contract for the local model policy training process on the edge server, used to record the training process and validate the locally trained model from each FL participant. The smart contract-based training policy will be validated prior to the aggregation process. Furthermore, we developed a smart contract-based aggregation policy for the global model aggregation process, recorded to capture important information for distributing the global model to each client. Afterward, all clients participating in the federated learning round receive the global model via the blockchain. The contributions of our work are summarized as follows: 

\begin{itemize}
\item We designed a verifiable and auditable management system for enhancing trustworthiness in the federated learning setting. 
\item We proposed a smart contract-based local training policy mechanism to ensure the training process is done correctly on the participant side and provide local model verifiability.
 \item We presented a smart contract-based global model aggregation policy to maintain global model integrity and provide participants with global model verifiability and global model access management. 
\end{itemize}

The rest of this paper is organized as follows. Section \ref{sec:issue} defines the problem. Section \ref{sec:related} discusses the related work. Then, we present the system architecture and introduce the proposed frameworks in Section \ref{sec:framework}. Next, we describe the proposed work's experimental setup and evaluation results in Section \ref{sec:exp}. Finally, a conclusion is drawn in Section \ref{sec:con}.

\section{Problem Scenario}\label{sec:issue}

\begin{figure*}[!h]
\centering
\includegraphics[width=0.9\linewidth]{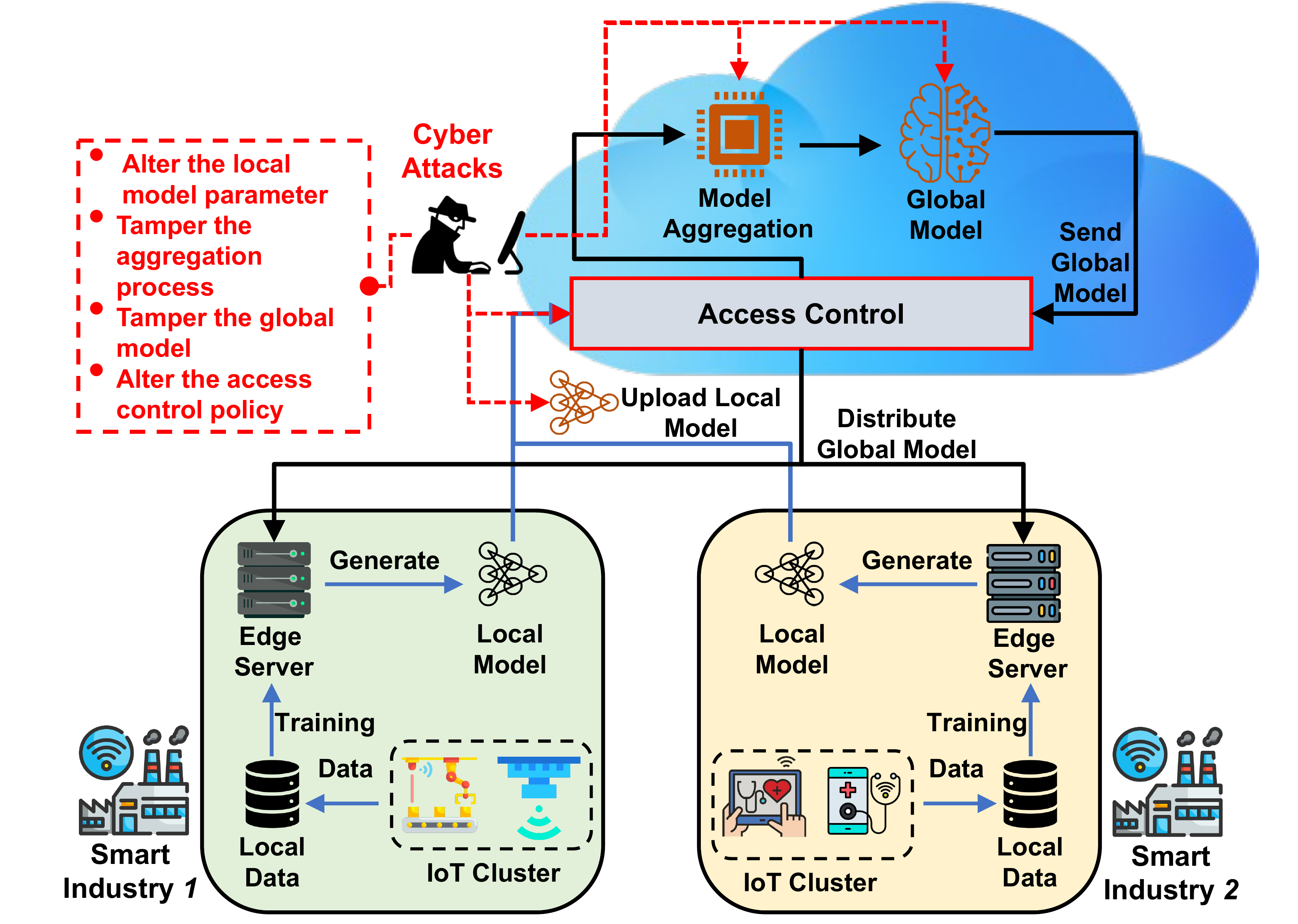}
\caption{\small{Possible threat on the current federated learning architecture}}
\label{fig:problem}
\end{figure*}

We use an IoT-based industry scenario to explore and highlight the existing challenges of current federated learning. We analyze the implications of these challenges with respect to the accuracy of machine learning models, data privacy, and security. In this context, assume that multiple smart manufacturers are located in different areas, each equipped with a set of IoT devices with sensors that capture and generate image data. As the IoT sensors are resource-constrained devices and cannot execute any machine learning algorithms, each manufacturer has an edge server as computing resources to perform machine learning tasks with their local datasets. However, the generated machine learning models have only moderate accuracy due to dataset limitations. Therefore, the edge servers from each manufacturer participate in a federated learning scenario to improve the accuracy of their models. In Federated Learning, local models from the manufacturer's edge server are gathered and aggregated to construct a highly precise machine learning model without sending any local datasets to the aggregation server. Afterward, the global model is sent back to the edge server for another round of federated learning. The global model is then used to recognize objects with greater accuracy upon achieving the desired accuracy.

Although Federated Learning has been demonstrated to improve machine learning accuracy, it is still susceptible to various security risks. Fig. \ref{fig:problem} illustrates the possible threat of the current FL architecture, such as:

\begin{itemize}
\item \textit{Risks of a faulty local model}:
In the current federated learning setup, each participant sends their local model to the cloud for the aggregation process. However, the aggregation server receives the local model without verifying the given local model, raising the risk of a local model being altered or poisoned. For example, an attacker can alter the local model parameters, causing a faulty local model. Unfortunately, the current FL architecture does not check whether the participants did the training process properly. Thus, validating the local model is essential to protect it from various security risks.

\item \textit{Risks of generating a biased aggregated model}: In the current federated learning global model aggregation protocol, the aggregation server receives local models from each participant and performs aggregation to generate the global model. However, this process can be easily tampered with, leading to a biased global model. For instance, an attacker can include a poisoned local model during the aggregation process, resulting in a false classification of the global model. Furthermore, the server does not verify the received local models, which can lead to a faulty local model and disrupt the entire aggregation process. To tackle this security risk, secure distributed aggregation and verifiable local models are needed.

\item \textit{Risk of receiving faulty global model}: In the existing federated learning method, the global model developed in the cloud is sent back to each edge server on the participants' side. However, the participants cannot verify the global model they receive, making it vulnerable to interception and alteration by malicious attackers. This can lead to the manufacturer receiving a faulty global model, necessitating the need for a global model verification method to ensure the integrity of the global model.

\end{itemize}

\section{Related Work}\label{sec:related}

Recently, researchers have proposed several studies to enhance the state-of-the-art federated learning architecture. In \cite{9352033}, the authors provide a comprehensive survey of the challenges and research directions of federated learning. Specifically, they discuss a range of topics, including management, security, privacy, scalability, and blockchain, to improve the current FL architecture. 

Trusted Execution Environments (TEEs) have emerged as a promising solution for preserving the privacy of machine learning models. In \cite{ohrimenko2016oblivious}, the authors investigate the use of SGX-enabled servers for machine learning to enhance data privacy and provide verifiability. Moreover, other works in \cite{9708971}, and \cite{juvekar2018gazelle} leverage TEEs to perform the aggregation process in a federated learning scenario, thus improving the security of the federated learning, albeit with increased time and computing power consumption.

Blockchain was initially developed for cryptocurrency purposes \cite{nakamoto2008bitcoin}. Since blockchain can maintain data integrity, it has evolved to enable distributed data storage across numerous computational nodes \cite{9617624}. Combining blockchain with federated learning (FL) can ensure the integrity of the machine learning model. Authors in \cite{9831779} and \cite{ali2021security} proposed a method to guarantee the privacy and security of a system using blockchain. Their method uses smart contracts and encryption to protect patient data from collision attacks. As the original federated learning architecture relies on a centralized server, researchers now leverage the blockchain in their federated learning methodology to secure the system. In \cite{9686048} and \cite{9930843}, a blockchain-based federated learning architecture is proposed, where each participant stores their locally trained model in the shared blockchain. However, since the current privacy measures do not protect the local models, other participants can gain access to them, raising serious privacy concerns. In \cite{9420107} and \cite{9822973}, the author proposed a blockchain-based FL healthcare scenario in which the local model is sent from the blockchain. However, model aggregation is done on a single server, and there are no verification processes in place before the aggregation, leaving the system vulnerable to a single point of failure and tampering attack. To address this, \cite{9170559} designed a verifiable local model using a multi-signature scheme. Each FL participant must sign the model for each FL round, and the cloud verifies each model for the aggregation process. However, the computation cost of the multi-signature scheme can be high when many clients join the FL round. To overcome this, \cite{9285303} proposed a verifiable aggregation for FL, which follows the idea of blockchain and uses a hash to calculate the digest for validation. Nonetheless, the aggregation and hashing operations are performed on a centralized server.

The work in \cite{9321132} leverages smart contracts to verify the integrity of the global model stored on the blockchain. In the initial round of federated learning, each FL participant receives the global model with a smart contract. The smart contract is executed in the participant's edge server to verify that the initial model has not been tampered. However, the proposed approach's smart contract is only utilized to verify the model's integrity; it has no information regarding the training and aggregation process during the model's development.

Author in \cite{10035023} and \cite{9234516} proposes a novel privacy framework for off-chain Federated Learning (FL), which incorporates blockchain and smart contracts for on-chain FL. The framework comprises private P2P identification and private FL modules, managed with scalable smart contracts to facilitate distributed collaborative mining with dynamic quantitative incentives. Furthermore, the framework utilizes a diffuse verified model to build an AI market with natural auditability and traceability. However, the proposed smart contract is fairly complex and inefficient due to the high deployment cost associated with it.

The paper in \cite{9235494} proposes an access control model for medical records in IoT-enabled smart healthcare devices using blockchain-based smart contracts. The scheme utilizes smart contracts to avoid network congestion and employs cryptographic functions for secure registration and retrieval of electronic medical records (EMR). The proposed model is implemented on the Ethereum private blockchain network and has been demonstrated to be a feasible solution for secure decentralized access control. However, the smart contract deployment cost is relatively high, and it is only used for access control.

After reviewing the aforementioned studies, due to obvious deficiencies of the prevalent approaches, this paper develops a smart contract-based policy control to manage the local model training and global model aggregation process in Federated Learning (FL) participants and to verify the integrity of the model. The smart contract-based policy control will record the core information during the training and aggregation to guarantee the model's integrity while providing an additional layer of access control to enhance the security of the FL architecture. This policy control will provide a secure and reliable approach to maintaining the privacy of the data and integrity of the model within the FL system.

\section{Proposed Framework}\label{sec:framework}
This section presents the proposed smart policy to control the federated learning management system. First, we present an overview of the system architecture. Next, we discuss in detail the various components of the proposed method. A summary of the notations used throughout the methodology is provided in Table I.

\begin{table}
\begin{center}
\caption{Notations}
\scalebox{1}{
\begin{tabularx}{\linewidth}{@{}XX@{}}
\toprule
    $P$ & Smart Factory\\
    $L_{n}$ & Local Image Dataset\\
    $Z_{n}$ & IoT Sensors\\
    $C_{n}$ & IoT Cluster\\
    $E_{n}$ & Edge Server\\
    $LM_{n}$ & Local Model \\
    $GM_{n}$ & Global Model \\
    $LM_{n}^{r+1}$ & {Updated Local Model}\\
    $GM_{n}^{r+1}$ & {Updated Global Model}\\
    $BA_{n}$ & Blockchain Aggregation Node\\
    $BD_{n}$ & Blockchain Database Node\\
    $PLM_{n}$ & Local Training Policy\\
    $PGM_{n}$ & Aggregation Policy\\
    $CSP$ & Cloud Service Provider\\
    $PCS$ & Policy Control Management System\\
    $BAM$ & Blockchain Aggregation Manager\\
    $BDM$ & Blockchain Database Manager\\

\bottomrule
\end{tabularx}
}
\end{center}
\end{table}

\subsection{System Architecture}

We proposed a secure management system for federated learning, leveraging smart contracts as the policy control for generating local models and aggregating them to generate a global model.

\begin{figure*}[!h]
\centering
\includegraphics[width=1\linewidth]{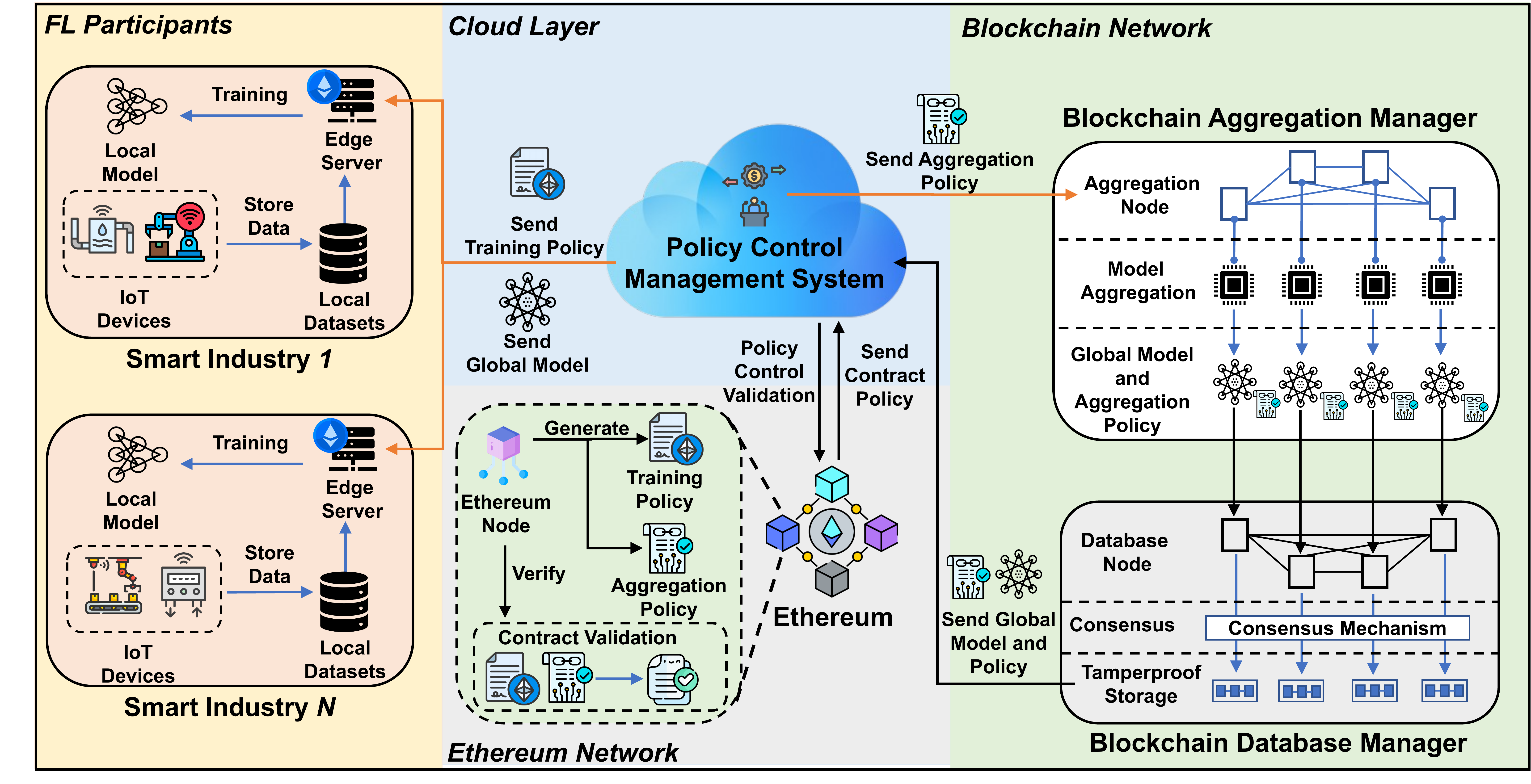}
\caption{\small{Overview of the proposed framework}}
\label{fig:architecture}
\end{figure*}

We assume that there is $P$ number of smart factories, each equipped with several IoT sensors $Z_n$, as a data source. Since IoT devices have limited computing resources, each smart factory has an edge server $E_n$ to support the computing process within the factory. Each edge server $E_n$ runs an Ethereum node connected to the Ethereum network, which allows it to execute the smart contract-based policies. The edge server can pre-process the data from the $Z_n$ and perform training for the machine learning model. As a result, each smart factory forms a cluster $C_n (1 \leq n \leq P)$. In the centralized machine learning approach, each smart factory performs the training process using its own local dataset $L_n$ produced by $Z_n$. However, due to the limited dataset of the machine learning model produced by the smart factory, the accuracy of the model may not be high enough to reach the desired level of precision and accuracy.
To address this limitation, the decentralized machine learning approach can be implemented by allowing the edge servers to collect and share data with other edge servers in the same cluster. By combining the data from different smart factories, the machine learning model can be trained with a larger and more diverse dataset, thus increasing its accuracy and reliability.

In this scenario, each smart factory joins the federated learning (FL) process in order to generate a high-accuracy model while maintaining the privacy of their respective local image dataset $L_n$. To generate the local model $LM_n$, each smart factory $P$ uses their edge server $E_n$. Since the FL process requires multiple participants with a dispersed range of datasets and policies, a smart contract-based local training policy $PLM_n$ is needed to ensure uniformity of the training process on the sides of the participants. The policy control management system ($PCS$) of the cloud service provider ($CSP$) will validate each of the $PLM_n$ with the Ethereum network before sending it to the blockchain manager $BM$ for the aggregation process. The final step in the FL process is to collect all the $LM_n$ and aggregate them into the global model $GM_n$.

However, in the original FL mechanism, participants do not know which parties join the FL process and contribute to the $GM_n$. To provide an auditable FL scheme, the Blockchain Aggregation Manager ($BAM$) will manage the Smart Contract Aggregation Policy ($PGM_n$). This policy will record the list of each $LM_n$ that participated in the current $GM_n$, and perform the aggregation process. After $GM_n$ is generated, it is stored concurrently with $PGM_n$ in the Blockchain Database Manager's (BDM) blockchain node. The $PCS$ will then distribute the $GM_n$ to each $P$ based on the $PGM_n$. By utilizing smart contracts and blockchain, each $P$ can verify the integrity of $GM_n$ and audit the $GM_n$. Fig. \ref{fig:architecture} provides an overview of the proposed framework. To make it easier to explain in later sections, the components of the proposed framework are broken down as follows:

\begin{itemize}
    \item {\textbf{Cloud Service Provider (CSP)} 
   acts as an intermediary between clients and blockchain, sending the contract policy and machine learning model to the clients and facilitating direct communication.}
    
    \item {\textbf{Policy Control Management System (PCS)}
    is hosted by the CSP and is responsible for verifying and managing the distribution of the $LM_n$ and $GM_n$ to the participants. To ensure trustworthiness, PCS communicates with the blockchain network to verify the integrity of the machine learning model.}
    
    \item {\textbf{Blockchain Aggregation Manager (BAM)} 
    communicates with PCS on behalf of the blockchain and receives the validated machine learning model from PCS. It then registers the participated machine learning model in $PGM_n$.}
    
    \item {\textbf{Blockchain Database Manager (BDM)}
    is a node on the blockchain network that is responsible for storing the $GM_n$ and $PGM_n$. The PCS  communicates with the BDM to request authenticated $GM_n$ and $PGM_n$ to manage the client's access to validated $GM_n$.}
\end{itemize}

\subsection{Smart Local Model Generation Policy for Local Model Generation}

A smart local model generation policy is performed by each participant to generate a locally-trained model based on the procedure provided by the PCS. In this method, each edge server $E_n$ receives the smart contract $PLM_n$ from the PCS. The purpose of this smart contract $PLM_n$ is to guarantee that all participants in the current FL round execute the training as per the given policy. The fields of the smart contract $PLM_n$ include the following information:

\begin{itemize}
    \item {\textit{ClientID}: This field records the unique identity of the edge server $E_n$ from each federated learning participant. The \textit{ClientID} is later used to retrieve the global model and to record the generation of the local model.}
    
    \item {\textit{ModelArchitecture}: these fields record the machine learning architecture used by the participants while performing the local model training process. The machine learning architecture needs to be recorded as the model aggregation process requires the same machine learning architecture.}
    
    \item {\textit{TrainingRound}: This field records the iteration of the federated learning rounds. The value will be used by $PCS$ to distribute the $GM_n$ since only the participants that join the current round of FL can obtain the aggregated model $GM_n$.}
    
    \item {\textit{Epoch}: The fields record the number of epoch when $E_n$ perform the local training. To validate the local model, $PCS$ will compare the number of epochs that are stored in the \textit{Epoch} in $PLM_n$ with the value that is recommended from the $PCS$. The value also can be used for further analysis by comparing the epoch with the accuracy.}
    
    \item {\textit{ModelAccuracy}: This field records the accuracy of the local model produced by each $E_n$. The value can be presented while performing the model aggregation in the blockchain.}
    
    \item {\textit{LocalModelHash}: The fields record the local model hash generated by the $E_n$. After $PCS$ receive the model from $E_n$, $PCS$ will compute the hash value of the $LM_n$ and compare the hash value with the \textit{LocalModelHash} that is stored in the $PLM_n$.}
    
\end{itemize}

\begin{figure}[!h]
\centering
\includegraphics[width=1\linewidth]{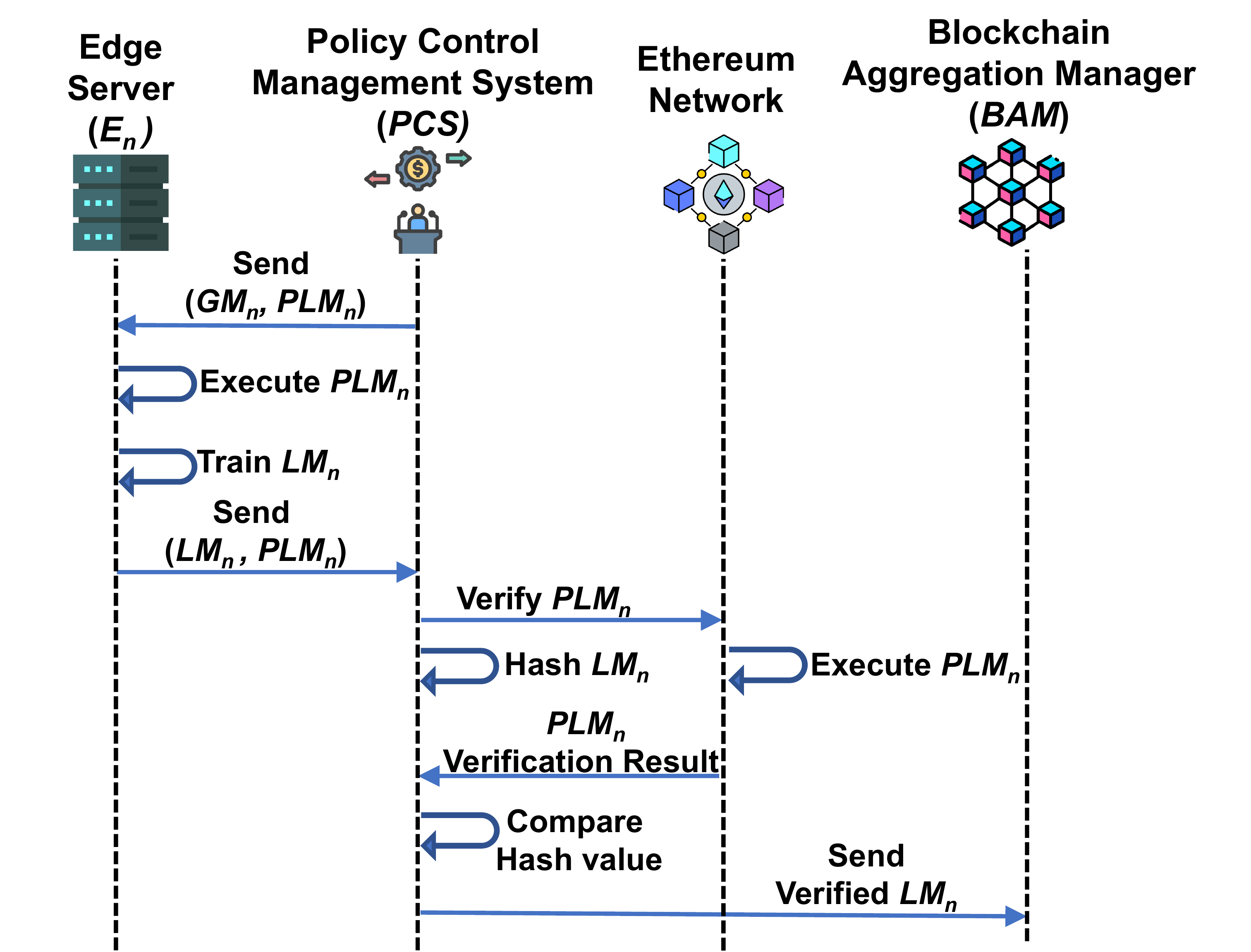}
\caption{\small{Workflow of the smart contract-based local policy training and verification process}}
\label{fig:local_model_gen}
\end{figure}

In the proposed system, $PCS$ sends the machine learning model for generating $LM_n$ and the $PLM_n$ for the policy to each $E_n$. Each $E_n$ executes the $PLM_n$ before the training process. We assume that edge servers of different $C_n$ train the models given from $PCS$ with Convolutional Neural Network (CNN)-based image classification. $PCS$ retrieves the initial machine learning model from $BDM$. The example of the CNN models are LeNet\cite{lecun1998gradient}, and AlexNet\cite{krizhevsky2017imagenet}. An overview of the smart local model generation policy phase is given in Fig. \ref{fig:local_model_gen}. 

In general, CNN classification performs prepossessing to an input image and classifies it under certain categories of objects. Edge server $E_n$ from the cluster $C_n$ generates a local dataset $L_n$. $E_n$ identifies the input image as an array of pixels based on the image resolution. Edge server read the image based on the $h \times w \times d$ ($h$ = Height, $w$ = Width, $d$ = Dimension). The CNN learning model uses different unique layers to train and test the local model. The layers used by the CNN model included  \textit{kernels}, \textit{pooling}, and \textit{fully connected layers}. At last, the CNN employs \textit{softmax function} to classify the given object according to the probability value among $0$ and $1$. In the FL scenario, every local model $LM_n$ is trained on the edge server $E_n$. Before starting the training process, $E_n$ execute the $PLM_n$. At first, $PLM_n$ record the \textit{ClientID} and initialize the training policy consisting of \textit{ModelArchitecture}, \textit{TrainingRound}, and \textit{Epoch}. After the training policy is executed, the edge server carries out the training utilizing its local dataset in every round $r$ of FL as follows:

\begin{equation}
  LM_{n}^{r+1} = GM_n^r - \eta \nabla F(GM_n^r,D^i) 
\end{equation}

Where $LM_{n}^{r+1}$ denotes the updated local model of client $p$, $GM_n^r$ is the current global model, $\eta$ is the local learning rate, $\nabla$ is used to refer to the derivative for every parameter, and $F$ is the loss function. After the local training process is finished, $E_n$ calculate the hash of the updated local model $LM_{n}^{r+1}$. Afterward, $PLM_n$ stores the \textit{ModelAccuracy} and the \textit{LocalModelHash}. Afterwards, each client will send $PLM_n$ and $LM_{n}^{r+1}$ to $PCS$ before forwarding them to the $BAM$ for the aggregation process.

Before delivering $LM_n$ to $PCS$, $E_n$ uses a symmetric key encryption algorithm, like Advanced Encryption Standard (AES), to encrypt the local model to ensure the security of the local model. We presume that the Diffie-Hellman key exchange mechanism or another secure key establishment mechanism was used to create the AES secret key between $E_n$ and the policy control management system. Algorithm \ref{alg:local_training} shows the step of smart contract-based policy training in detail.

\begin{algorithm}
\SetAlgoLined
\KwIn{

Initial global model $GM_n$

Training policy $PLM_n$

Local dataset $L_n$
}
\KwOut{

Updated local model $LM_{n}^{r+1}$

Updated training policy $PLM_n$

}

\While{Edge server $E_n$ is running}
{
\textbf{Execute} 

Training Policy $PLM_n$

\While{Training policy $PLM_n$ is running}
{
\textbf{Record:}

Integer \textit{ClientID}

\textbf{Training Policy:}

Integer \textit{ModelArchitecture}

Integer \textit{TrainingRound}\

Integer \textit{Epoch}\

\textbf{Initialize Training:}
Load dataset $L_n$\
\ForEach{epoch}{
\textbf{Shuffle} the training data $D$
\ForEach{each training sample $(x,y) \in L_n$}{
Calculate the gradient

Calculate the loss function

Update the model parameters

Local training, as shown in (1)
}
}
Hash the $LM_{n}^{r+1} \rightarrow H(LM_{n}^{r+1})$

\textbf{Training Policy Log:}

String \textit{ModelAccuracy} 

String \textit{LocalModelHash} $\leftarrow H(LM_{n}^{r+1})$
}
\textbf{Return}

($PLM_n$, $LM_{n}^{r+1}$)

Encrypt $LM_{n}^{r+1} \rightarrow E(LM_{n}^{r+1})$

\textbf{Send} ($PLM_n$, $E(LM_{n}^{r+1})$) to $PCS$

}
\caption{Smart contract-based training policy}
\label{alg:local_training}
\end{algorithm}

\subsection{Smart Policy Control Model Aggregation}

After the policy control management system verifies the authenticity of $LM_n$ by executing $PLM_n$ from each FL participant, $LM_n$ is sent to the blockchain aggregation manager for the aggregation process. Before generating the global model, $PCS$ generates a smart contract policy for the aggregation process $PGM_n$ and sends it to $BAM$. In our scenario, $PGM_n$ is used to ensure the global model is generated based on the trusted participants and maintain the security of the aggregation process. Later, $PGM_n$ is used by $PCS$ for distributing the $GM_n$ to the participants. The participant also receives $PGM_n$ to verify the authenticity of the $GM_n$. The fields of the smart contract $PGM_n$ contain the following information:

\begin{itemize}
    \item \textit{GlobalModelID}: These fields record the unique identity of the aggregated model. These fields are used for traceability and storing processes.
    
    \item \textit{FedRound}: These fields record the federated learning training round. This information is important for the auditing process for both servers and participants.
    
    \item \textit{ParticipantNum}: The fields is use to record the \textit{ClientID} that participate in the FL round. This field is used when $PCS$ distributes the $GM_n$ to the client. Hence, only \textit{ClientID} that participates in the training round receives the global model.
    
    \item \textit{GlobalModelAcc}: This field records $GM_n$ accuracy in the current federated learning round. This field is used for logging and fine-tuning the global model.
    
    \item \textit{GlobalModelHash}: This field records the hash of the current global model generated by \textit{BAM}. This information is used by \textit{BDM} to verify the authenticity of the global model before storing it on the blockchain.
    
\end{itemize}

We assume that there are $BA_n (1\leq n \leq b)$ blockchain nodes in the $BAM$. After the $PCS$ verifies the $LM_n$ with $PLM_n$, $PCS$ sends $LM_n$ and $PGM_n$ to $BAM$. $BAM$ sends a set of local models from all participants, which can be denoted as $LM = \{LM_{1}, LM_{2}, \hdots, LM_{n}\}$, and $PGM_n$, to each of the blockchain aggregation nodes $BA_{n}$. After each $BA_n$ receives the set of $LM$, each $BA_n$ executes $PGM_n$ to initialize the aggregation policy. Before the aggregation process starts, $PGM_n$ records the \textit{GlobalModelID}, \textit{FedRound}, and \textit{ParticipantNum}. Then, each $BA_n$ generates the aggregated global model $GM_n$ Federated Averaging (FedAVG) \cite{mcmahan2017communication} denoted as follows:

\begin{equation}
  GM_{n}^{r+1} = \sum_{n=1}^{p} \frac{|D_n|}{N} LM_{n}^{r+1}, N = \sum_{n=1}^{p} |D_n|  
\end{equation}

where $GM_{n}^{r+1}$ denotes the updated global model, $p$ is the number of clients on the federated learning round $r$, $|D_n|$ is the number of data items (images) owned by $E_{n}$ to train local model $LM_{n}^{r+1}$, and $N$ the total number of data used to train all of the local models. $GM_{n}$ is final updated global model $GM_{n}^{r+1}$. Since we leverage Federated Averaging (FedAVG) \cite{mcmahan2017communication} as the aggregation method, therefore, our proposed work is suitable for aggregating different types of neural network models, which can take various types of inputs (e.g., text and numerical values).

\begin{figure}[!h]
\centering
\includegraphics[width=1\linewidth]{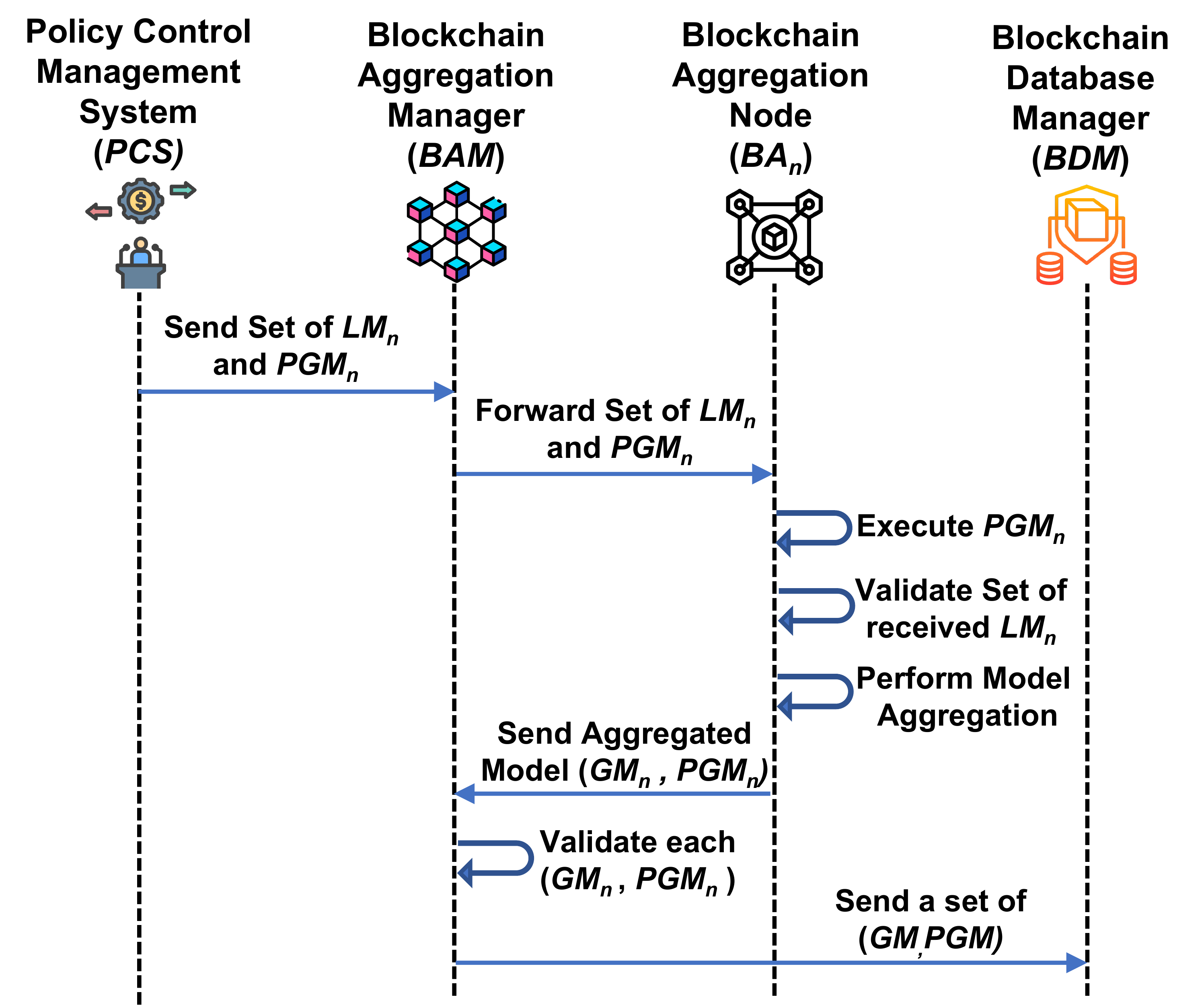}
\caption{\small{Workflow of smart contract-based aggregation policy for global Model Aggregation}}
\label{fig:aggregation}
\end{figure}

After the aggregation process finishes, each $BA_n$ hashes the updated global model $GM_{n}^{r+1}$ as the requirement of the $PGM_n$. Later $PGM_n$ records \textit{GlobalModelAcc}, and \textit{GlobalModelHash}. Afterward, $BAM$ validates each global model generated by $BA_n$ with the respective $PGM_n$. Once each of the $GM_{n}$ is validated, $BAM$ sends a set of $(GM,PGM) = \{(GM_{1},PGM_{1}), (GM_{2},PGM_{2}), \hdots, (GM_{n},PGM_{n})\}$ to blockchain database manager $BDM$. The overview of the smart policy aggregation is given in Fig. \ref{fig:aggregation}. Algorithm \ref{alg:aggregation} shows the steps of smart contract-based aggregation policy in detail.

\begin{algorithm}

\KwIn{

Locally trained models $LM^{n} = {LM_{1}, \ldots, LM_{n}}$

Local model policies $PLM^{n} = {PLM_{1}, \ldots, PLM_{n}}$

Aggregation policy $PGM_n$
}
\KwOut{

Aggregated global model $GM_{n}$

Updated aggregation policy $PGM_n$
}

\While{Aggregation server is running}
{
\textbf{Execute}

Aggregation Policy $PGM_n$

\While{Aggregation policy $PGM_n$ is running}
{

\textbf{Record:}

Integer \textit{GlobalModelID}

Integer \textit{FedRound}

Integer \textit{ParticipantNum}

{
\textbf{Initialize:}

Memory buffer, $Mem = \emptyset$

\ForEach{$LM_i \in LM_n$}
{
Check the hash of $LM_i$ with $PLM_i$

Add $LM_i$ to memory buffer $Mem$

}
Check criteria in $PGM_n$ for aggregation

\If{the set hash of $LM_n$ is all valid}
{
Aggregate all local models in $Mem$ using FedAvg algorithm as shown in (3) and generate global model $GM_{n}$
}

Hash $GM_{n} \rightarrow H(GM_{n})$

\textbf{Aggregation Policy Log:}

String \textit{GlobalModelAcc}

String \textit{GlobalModelHash} $\leftarrow H(GM_{n})$

Generate $PGM_n$ report

}
}
\textbf{return} $({GM_{n}, PGM_n})$
}
\caption{Smart Policy Global Model Aggregation Process}
\label{alg:aggregation}
\end{algorithm}

\subsection{Blockchain-based Tamperproof Storage}

\begin{figure}[!h]
\centering
\includegraphics[width=1\linewidth]{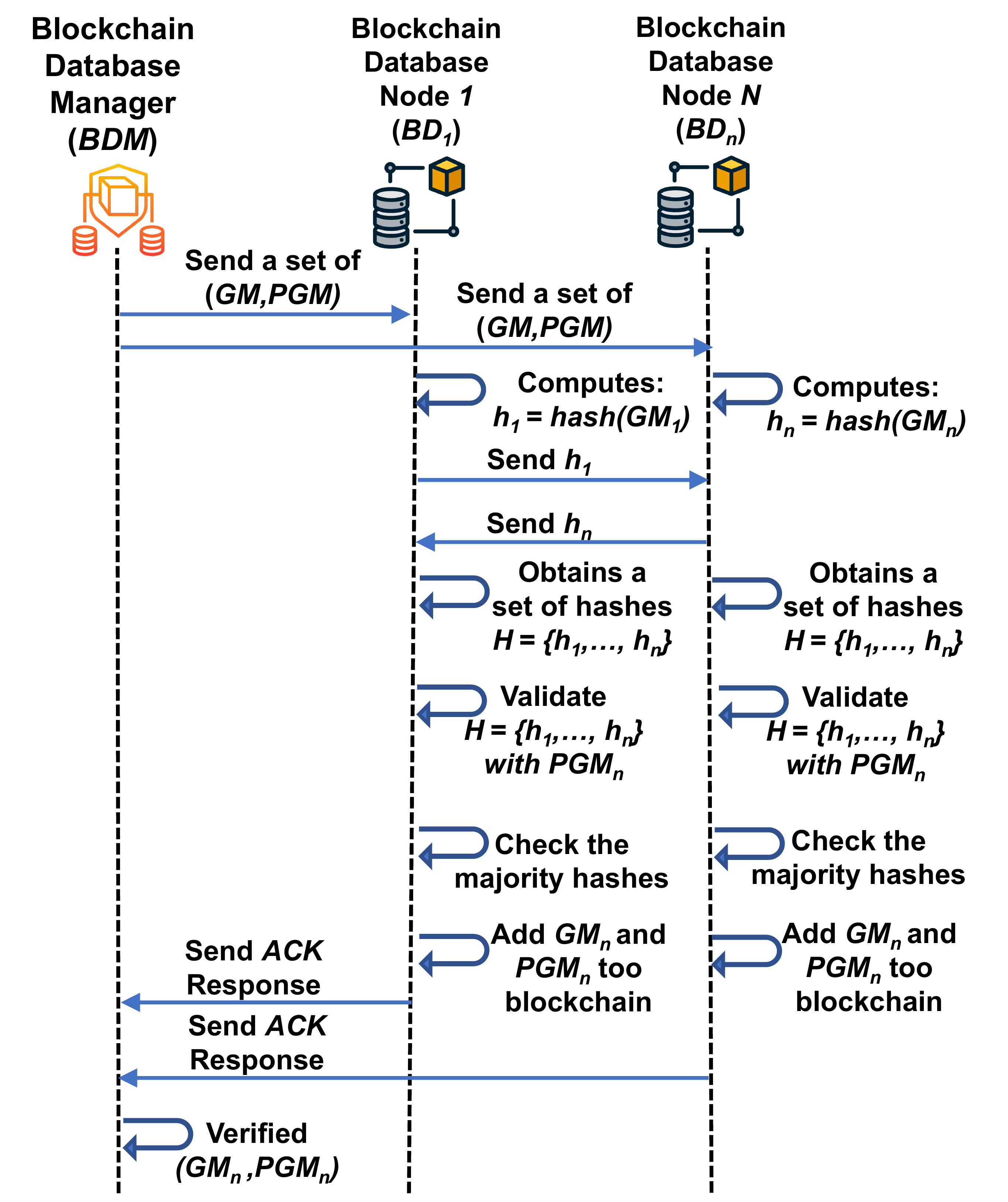}
\caption{\small{Workflow of the blockchain-based global model storage system}}
\label{fig:consensus}
\end{figure}

In this phase (see Fig. \ref{fig:consensus}), $BAM$ sends a set of $(GM,PGM)$ to $BDM$. $BDM$ needs to verify $GM_n$ by performing a consensus mechanism. The consensus mechanism verifies each $GM_n$ and $PGM_n$ produced by $BA_n$. If all the $GM_n$ and $PGM_n$ are verified, and the majority of the hashes of the corresponding $GM_n$ are the same, the blockchain nodes in $BDM$ add $(GM,PGM)$ as a block in the blockchain.

The consensus mechanism in $BDM$ has several steps to verify and stores the global model. At first, $BDM$ distributes the set of $(GM,PGM)$ to each $BD_n$. Later, each $BD_n$ calculates the hash of each $GM_n$ and compares it with the hash that is recorded in $PGM_n$. The consensus is achieved if the hashes of all $GM_n$ are the same. However, if all hashes are not the same, the blockchain node $BD_{n}$ in $BDM$ determines the global model that has the maximum matched hash values. Each $BD_{n}$ proposes $GM_n$ to the blockchain database manager $BDM$ to add to the blockchain. Finally, if $GM_n$ is the same for the majority of the node's global model, the consensus is achieved and added to the blockchain tamperproof storage. 

\begin{figure}[!h]
\centering
\includegraphics[width=1\linewidth]{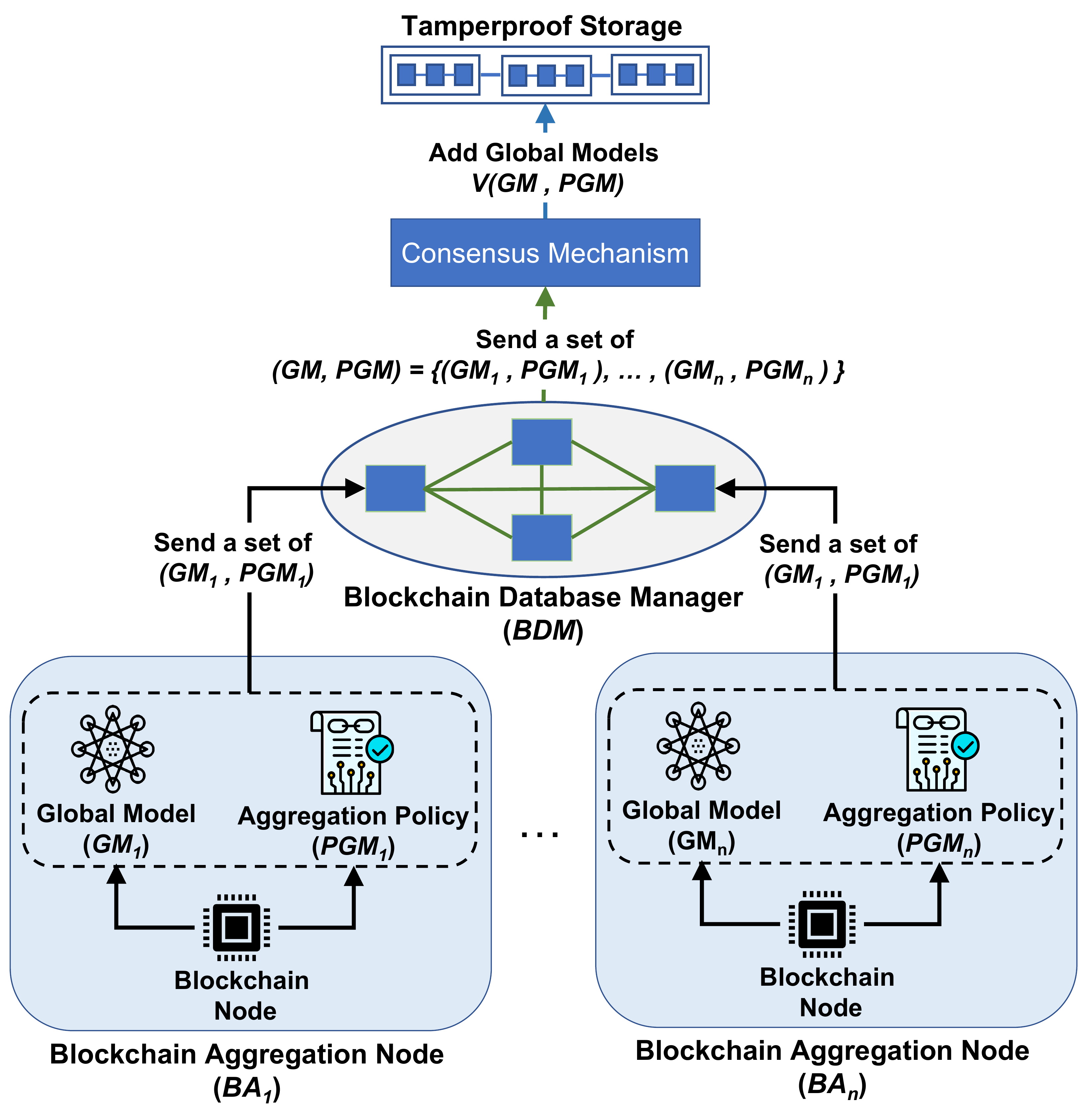}
\caption{\small{Storing global model on tamperproof storage}}
\label{fig:overviewconsensus}
\end{figure}

Fig. \ref{fig:overviewconsensus} provides an overview of the workflow for storing the global model on a blockchain-based, tamper-proof storage. Later, the global model $GM_{n}$ and $PGM_n$ are sent to $PCS$ for validation before $PCS$ sends the $GM_n$ to all edge servers $E_n$. The aggregated global model is distributed according to the \textit{ClientID} that is recorded in the smart policy contract.

\section{Results and Discussion}\label{sec:exp}
In this section, we discuss several experiments conducted to evaluate the performance of our proposed framework. Experimental setup and dataset and model are discussed in Section \ref{sec:setup} and \ref{sec:data}, respectively. Section \ref{sec:result} shows experimental results and evaluates the performance.

\subsection{Experimental Setup}\label{sec:setup}
In our experiments, we ran the experiment on the AWS EC2 cloud. To handle the local training process, which requires considerable computing power, we used the \textit{P3} machine instance \textit{ml.p3.8xlarge}. This machine had 4 NVIDIA Tesla V100 with 64 GB of memory, and a Peer-to-Peer connection between the GPUs, as well as 32 vCPUs and 244 GB of RAM. We built our federated learning application using PyTorch \cite{paszke2019pytorch}, and leveraged Ethereum, emulated in Ganache \cite{lee2019testing}, for the blockchain.

\subsection{Datasets and Machine Learning Architecture}\label{sec:data}

For the experiments, we selected two widely used datasets to benchmark the machine learning process: CIFAR-10 \cite{krizhevsky2010convolutional} and MNIST \cite{deng2012mnist}. These datasets are commonly used for evaluation in the machine learning framework. Thus, we utilized them to assess the performance of our proposed framework. The proposed method uses the dataset to train and test the local model on the client side. When performing the experiments, we split the training and test sets. We evenly distributed the training and test sets amongst the federated learning participants based on the number of clients. MNIST \cite{deng2012mnist} consists of 60,000 images in the training set and 10,000 in the test set of handwritten digits. Each image is a 28×28-pixel image of a handwritten digit. CIFAR-10 \cite{krizhevsky2010convolutional} consists of 50,000 images in the training set and 10,000 in the test set of 10 different classes (such as cars, dogs, and planes), and there are 6,000 images in each class, where each image contains 32×32-colored pixels. Table \ref{tab3} overviews the dataset used in the experiments.

\begin{table}[h]
\begin{center}
\resizebox{\columnwidth}{!}{
    \begin{tabular}{ |c|c|c|c|c| }
        \hline
        \textbf{Datasets} & \textbf{Training set} & \textbf{Test set} & \textbf{Size} & \textbf{Color}\\
        
        \hline
        MNIST \cite{deng2012mnist} & 60.000 & 10.000 & 28x28 & Grayscale\\
        \hline
        CIFAR-10 \cite{krizhevsky2010convolutional} & 50.000 & 10.000 & 32x32 & RGB\\
        \hline
    \end{tabular}
    }
    \end{center}
    \caption{Datasets specifications}
    \label{tab3}
\end{table}

We consider two machine learning architectures for our experiment: LeNet \cite{lecun1998gradient} and AlexNet \cite{krizhevsky2017imagenet}. LeNet has five layers consisting of two convolutional layers, two pooling layers, and one fully connected layer, while AlexNet has eight layers consisting of five convolutional layers and three fully-connected layers. AlexNet can also use batch normalization layers for stability and efficient training. In terms of parameters, LeNet has around 60,000, while AlexNet has around 60 million. We chose these two machine learning architectures to test our framework against learning models with diverse computational resources.

\subsection{Experimental Results and Performance Evaluation}\label{sec:result}

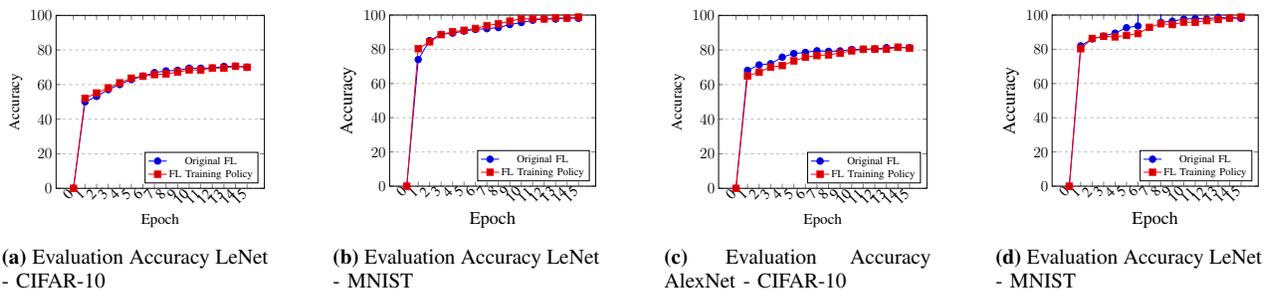
\begin{figure*}[tbh!]
\centering
\begin{subfigure}[tbh!]{0.4\columnwidth}
    \resizebox{1\columnwidth}{!}{
            \begin{tikzpicture}
                \begin{axis}[
                    xlabel={Epoch},
                    ylabel={Accuracy},
                    symbolic x coords = {0,1,2,3,4,5,6,7,8,9,10,11,12,13,14,15},
                    xticklabel style={anchor= east,rotate=45 },
                    xtick=data,
                    ymax=100,
                    ymin=0,
                    legend pos=south east,
                    ymajorgrids=true,
                    grid style=dashed,
                    legend style={nodes={scale=1, transform shape}},
                    label style={font=\Large},
                    tick label style={font=\Large}
                ]
                \addplot+[mark size=3pt]
                    coordinates {
                    (0,0)
                    (1,50.01)
                    (2,53.15)
                    (3,57.11)
                    (4,60)
                    (5,62.9)
                    (6,64.9)
                    (7,66.9)
                    (8,67.9)
                    (9,68.22)
                    (10,69.51)
                    (11,69.45)
                    (12,69.58)
                    (13,70.51)
                    (14,70.62)
                    (15,70.03)
                    };
                \addplot+[mark size=3pt]
                    coordinates {
                    (0,0)
                    (1,52.01)
                    (2,55.15)
                    (3,58.11)
                    (4,61)
                    (5,63.71)
                    (6,64.83)
                    (7,65.77)
                    (8,66.19)
                    (9,67.22)
                    (10,68.51)
                    (11,68.45)
                    (12,69.58)
                    (13,69.51)
                    (14,70.62)
                    (15,70.03)
                    };
                \legend{Original FL, FL Training Policy}
                \end{axis}
            \end{tikzpicture}
            }
    \caption{Evaluation Accuracy LeNet - CIFAR-10}
    \label{exp0a}
\end{subfigure}
~
~
~
\begin{subfigure}[tbh!]{0.4\columnwidth}
    \resizebox{1\columnwidth}{!}{
        \begin{tikzpicture}
                \begin{axis}[
                    xlabel={Epoch},
                    ylabel={Accuracy},
                    symbolic x coords = {0,1,2,3,4,5,6,7,8,9,10,11,12,13,14,15},
                    xticklabel style={anchor= east,rotate=45 },
                    xtick=data,
                    ymax=100,
                    ymin=0,
                    legend pos=south east,
                    ymajorgrids=true,
                    grid style=dashed,
                    legend style={nodes={scale=1, transform shape}},
                    label style={font=\LARGE},
                    tick label style={font=\Large}
                ]
                \addplot+[mark size=3pt]
                    coordinates {
                    (0,0)
                    (1,74)
                    (2,85.07)
                    (3,88.73)
                    (4,89.47)
                    (5,90.61)
                    (6,91.63)
                    (7,92.04)
                    (8,92.69)
                    (9,94.44)
                    (10,95.63)
                    (11,96.98)
                    (12,97.74)
                    (13,97.67)
                    (14,98.27)
                    (15,98.08)
                    };
                \addplot+[mark size=3pt]
                    coordinates {
                    (0,0)
                    (1,80.33)
                    (2,84.35)
                    (3,88.57)
                    (4,90.23)
                    (5,91.14)
                    (6,92.17)
                    (7,93.86)
                    (8,94.91)
                    (9,96.46)
                    (10,97.87)
                    (11,97.78)
                    (12,97.74)
                    (13,98.45)
                    (14,98.23)
                    (15,98.85)
                    };
                \legend{Original FL, FL Training Policy}
                \end{axis}
            \end{tikzpicture}
        }
    \caption{Evaluation Accuracy LeNet - MNIST}
    \label{exp0b}
\end{subfigure}
~
~
~
\begin{subfigure}[tbh!]{0.4\columnwidth}
    \resizebox{1\columnwidth}{!}{
            \begin{tikzpicture}
                \begin{axis}[
                    xlabel={Epoch},
                    ylabel={Accuracy},
                    symbolic x coords = {0,1,2,3,4,5,6,7,8,9,10,11,12,13,14,15},
                    xticklabel style={anchor= east,rotate=45 },
                    xtick=data,
                    ymax=100,
                    ymin=0,
                    legend pos=south east,
                    ymajorgrids=true,
                    grid style=dashed,
                    legend style={nodes={scale=1, transform shape}},
                    label style={font=\Large},
                    tick label style={font=\Large}
                ]
                \addplot+[mark size=3pt]
                    coordinates {
                    (0,0)
                    (1,68.23)
                    (2,71.36)
                    (3,72.05)
                    (4,75.86)
                    (5,77.94)
                    (6,78.60)
                    (7,79.53)
                    (8,79.19)
                    (9,79.46)
                    (10,80.30)
                    (11,80.35)
                    (12,80.57)
                    (13,81.33)
                    (14,81.50)
                    (15,81.33)
                    };
                \addplot+[mark size=3pt]
                    coordinates {
                    (0,0)
                    (1,65.01)
                    (2,67.15)
                    (3,70.11)
                    (4,71)
                    (5,73.71)
                    (6,75.83)
                    (7,76.77)
                    (8,77.19)
                    (9,78.22)
                    (10,79.51)
                    (11,80.45)
                    (12,80.58)
                    (13,80.51)
                    (14,81.62)
                    (15,81.03)
                    };
                \legend{Original FL, FL Training Policy}
                \end{axis}
            \end{tikzpicture}
            }
    \caption{Evaluation Accuracy AlexNet - CIFAR-10}
    \label{exp0c}
\end{subfigure}
~
~
~
\begin{subfigure}[tbh!]{0.4\columnwidth}
    \resizebox{1\columnwidth}{!}{
        \begin{tikzpicture}
                \begin{axis}[
                    xlabel={Epoch},
                    ylabel={Accuracy},
                    symbolic x coords = {0,1,2,3,4,5,6,7,8,9,10,11,12,13,14,15},
                    xticklabel style={anchor= east,rotate=45 },
                    xtick=data,
                    ymax=100,
                    ymin=0,
                    legend pos=south east,
                    ymajorgrids=true,
                    grid style=dashed,
                    legend style={nodes={scale=1, transform shape}},
                    label style={font=\LARGE},
                    tick label style={font=\Large}
                ]
                \addplot+[mark size=3pt]
                    coordinates {
                    (0,0)
                    (1,82)
                    (2,86.07)
                    (3,87.73)
                    (4,89.47)
                    (5,92.61)
                    (6,93.63)
                    (7,894.04)
                    (8,95.69)
                    (9,96.44)
                    (10,97.63)
                    (11,97.98)
                    (12,97.74)
                    (13,98.67)
                    (14,98.27)
                    (15,98.08)
                    };
                \addplot+[mark size=3pt]
                    coordinates {
                    (0,0)
                    (1,80.33)
                    (2,86.35)
                    (3,87.57)
                    (4,87.23)
                    (5,88.14)
                    (6,89.17)
                    (7,92.86)
                    (8,94.91)
                    (9,94.46)
                    (10,95.87)
                    (11,95.78)
                    (12,96.74)
                    (13,97.45)
                    (14,98.23)
                    (15,98.85)
                    };
                \legend{Original FL, FL Training Policy}
                \end{axis}
            \end{tikzpicture}
        }
    \caption{Evaluation Accuracy LeNet - MNIST}
    \label{exp0d}
\end{subfigure}
\caption{Federated learning evaluation accuracy with and without training policy (a)with LeNet and CIFAR-10 datasets; (b)with LeNet and MNIST datasets; (c)with AlexNet Model and CIFAR-10 datasets; (d)with AlexNet Model and MNIST datasets}
\label{exp0}
\end{figure*}

In Fig. \ref{exp0}, we compare the evaluation accuracy of machine learning architectures and datasets with and without the training policy. Figs. \ref{exp0a} and \ref{exp0c} show the effect of the smart contract-based training policy compared to the original FL architecture on the accuracy of LeNet and AlexNet when using CIFAR-10 datasets. The peak evaluation accuracy for both methods is 70\% when applied to CIFAR-10 datasets. When using AlexNet architecture, both methods can reach 80\%. This is because LeNet architecture is smaller, with up to sixty million parameters. Figs. \ref{exp0b} and \ref{exp0d} show the effect of the smart contract-based training policy compared to the original FL architecture on the accuracy of LeNet and AlexNet when using MNIST datasets. The evaluation accuracy from both architectures can get up to 90\%. With MNIST datasets, LeNet architecture can keep up with AlexNet since MNIST datasets are relatively simple and not as complex as CIFAR-10 datasets. The result from Fig. \ref{exp0} shows that our smart contract-based training policy did not affect the accuracy of the machine learning.

\begin{figure*}[tbh!]
\centering
\begin{subfigure}[tbh!]{0.4\columnwidth}
    \resizebox{1\columnwidth}{!}
    {
        \begin{tikzpicture}
                    \begin{axis}[
                    xlabel={Number of Clients},
                    ylabel={Processing Time (Min)},
                    symbolic x coords = {2,5,10,15,20},
                    xticklabel style={anchor= east,rotate=45 },
                    xtick=data,
                    ymax=10,
                    ymin=0,
                    legend pos=south east,
                    ymajorgrids=true,
                    grid style=dashed,
                    legend style={nodes={scale=1, transform shape}},
                    label style={font=\Large},
                    tick label style={font=\Large}
                ]
                \addplot+[mark size=3pt]
                    coordinates {
                        (2,3.05)
                        (5,3.49)
                        (10,4.35)
                        (15,5.42)
                        (20,6.86)
                    };
                \addplot+[mark size=3pt]
                    coordinates {
                        (2,3.30)
                        (5,4.2)
                        (10,5.05)
                        (15,6.05)
                        (20,7.91)
                    };
                \legend{Original FL, Fl Training Policy}
                \end{axis}
            \end{tikzpicture}      
        }
    \caption{LeNet - MNIST}
    \label{exp1a}
\end{subfigure}
~
~
~
\begin{subfigure}[tbh!]{0.4\columnwidth}
    \resizebox{1\columnwidth}{!}{
    \begin{tikzpicture}
                \begin{axis}[
                    xlabel={Number of Clients},
                    ylabel={Processing Time (Min)},
                    symbolic x coords = {2,5,10,15,20},
                    xticklabel style={anchor= east,rotate=45 },
                    xtick=data,
                    ymax=25,
                    ymin=0,
                    legend pos=south east,
                    ymajorgrids=true,
                    grid style=dashed,
                    legend style={nodes={scale=1, transform shape}},
                    label style={font=\Large},
                    tick label style={font=\Large}
                ]
                \addplot+[mark size=3pt]
                    coordinates {
                        (2,12.2)
                        (5,14.5)
                        (10,16.05)
                        (15,18.85)
                        (20,19.11)
                    };
                \addplot+[mark size=3pt]
                    coordinates {
                        (2,13.2)
                        (5,16.5)
                        (10,18.05)
                        (15,20.85)
                        (20,22.11)
                    };
                \legend{Original FL, Fl Training Policy}
                \end{axis}
            \end{tikzpicture}
        }
    \caption{LeNet - CIFAR-10}
    \label{exp1b}
\end{subfigure}
~
~
~
\begin{subfigure}[tbh!]{0.4\columnwidth}
    \resizebox{1\columnwidth}{!}
    {
    \begin{tikzpicture}
                \begin{axis}[
                    xlabel={Number of Clients},
                    ylabel={Processing Time (Min)},
                    symbolic x coords = {2,5,10,15,20},
                    xticklabel style={anchor= east,rotate=45 },
                    xtick=data,
                    ymax=10,
                    ymin=0,
                    legend pos=south east,
                    ymajorgrids=true,
                    grid style=dashed,
                    legend style={nodes={scale=1, transform shape}},
                    label style={font=\Large},
                    tick label style={font=\Large}
                ]
                \addplot+[mark size=3pt]
                    coordinates {
                        (2,5.06)
                        (5,5.12)
                        (10,6.22)
                        (15,7.05)
                        (20,7.9)
                    };
                \addplot+[mark size=3pt]
                    coordinates {
                        (2,5.62)
                        (5,5.95)
                        (10,7.05)
                        (15,7.85)
                        (20,8.81)
                    };
                \legend{Original FL, Fl Training Policy}
                \end{axis}
            \end{tikzpicture}
    }
    \caption{AlexNet - MNIST}
    \label{exp1c}
\end{subfigure}
~
~
~
\begin{subfigure}[tbh!]{0.4\columnwidth}
    \resizebox{\columnwidth}{!}
    {
    \begin{tikzpicture}
                \begin{axis}[
                    xlabel={Number of Clients},
                    ylabel={Processing Time (Min)},
                    symbolic x coords = {2,5,10,15,20},
                    xticklabel style={anchor= east,rotate=45 },
                    xtick=data,
                    ymax=30,
                    ymin=0,
                    legend pos=south east,
                    ymajorgrids=true,
                    grid style=dashed,
                    legend style={nodes={scale=1, transform shape}},
                    label style={font=\Large},
                    tick label style={font=\Large}
                ]
                \addplot+[mark size=3pt]
                    coordinates {
                        (2,13.2)
                        (5,16.5)
                        (10,18.05)
                        (15,20.85)
                        (20,22.11)
                    };
                \addplot+[mark size=3pt]
                    coordinates {
                        (2,15.2)
                        (5,18.5)
                        (10,20.05)
                        (15,23.85)
                        (20,26.11)

                    };
                \legend{Original FL, Fl Training Policy}
                \end{axis}
            \end{tikzpicture}
    }
    \caption{AlexNet - CIFAR 10}
    \label{exp1d}
\end{subfigure}
\caption{Processing time of training process with and without training policy using various machine learning models and datasets.}
\label{exp1}
\end{figure*}
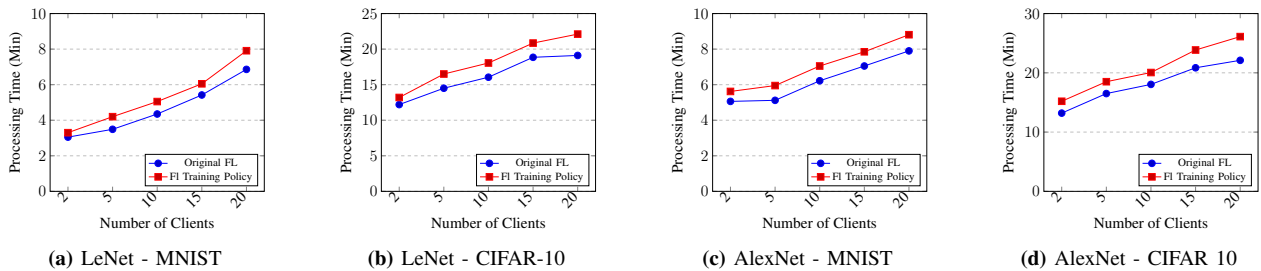

In Fig. \ref{exp1}, we evaluate the performance of our proposed framework for the local model training process. This experiment reveals the local model training time cost difference between the original federated learning setup and using the smart policy local training process. To perform the comparison, we run one round of federated learning using default settings, with clients' nodes ranging from two to twenty clients. We then perform another round of federated learning by enabling the local training policy. In this experiment, we concurrently perform the federated learning process to observe the effect of the training policy on the training process. Results show the local model training time cost required by LeNet and AlexNet using MNIST and CIFAR-10 datasets. 

In Figures \ref{exp1a} and \ref{exp1b}, the results of the LeNet model when performing local training using regular FL and policy-based training using MNIST and CIFAR-10 datasets are shown. The time cost is consistently stable from two to five clients but begins to increase gradually when there are ten to twenty clients. When using the LeNet learning model in a policy-based training setup, the average time cost increases by 1-2 minutes compared to the original FL setup. Figures \ref{exp1c} and \ref{exp1d} compare the AlexNet model using MNIST and CIFAR-10 datasets. Compared to LeNet, the overall time of AlexNet is higher due to its larger number of layers. The time cost in the AlexNet model steadily increases when it has five clients. The time cost when performing federated learning with a local model training policy also increases slightly compared to the original FL setup. The experimental results from both models demonstrate that the time cost increases linearly for both the original FL training and policy-based FL training. Policy-based local training is slightly higher than the original one since every FL participant needs to execute the training policy contract before the training process commences.

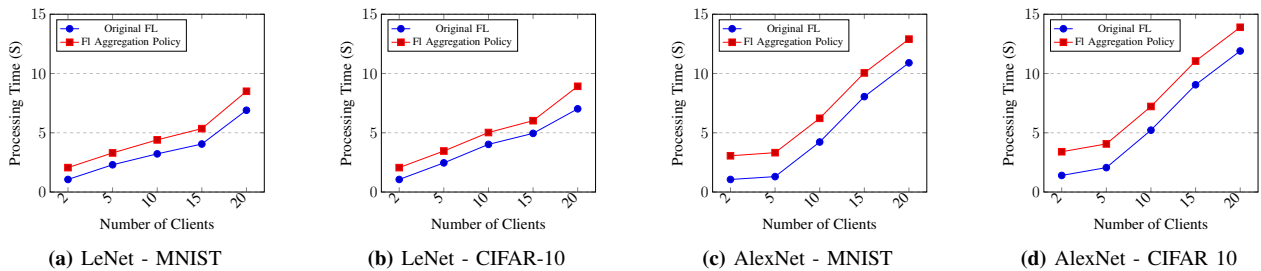
\begin{figure*}[tbh!]
\centering
\begin{subfigure}[tbh!]{0.4\columnwidth}
    \resizebox{1\columnwidth}{!}
    {
        \begin{tikzpicture}
                \begin{axis}[
                    xlabel={Number of Clients},
                    ylabel={Processing Time (S)},
                    symbolic x coords = {2,5,10,15,20},
                    xticklabel style={anchor= east,rotate=45 },
                    xtick=data,
                    ymax=15,
                    ymin=0,
                    legend pos=north west,
                    ymajorgrids=true,
                    grid style=dashed,
                    legend style={nodes={scale=1, transform shape}},
                    label style={font=\Large},
                    tick label style={font=\Large}
                ]
                \addplot+[mark size=3pt]
                    coordinates {
                        (2,1.06)
                        (5,2.30)
                        (10,3.22)
                        (15,4.05)
                        (20,6.9)
                    };
                \addplot+[mark size=3pt]
                    coordinates {
                        (2,2.06)
                        (5,3.30)
                        (10,4.40)
                        (15,5.35)
                        (20,8.5)
                    };
                \legend{Original FL, Fl Aggregation Policy}
                \end{axis}
        \end{tikzpicture}
        }
    \caption{LeNet - MNIST}
    \label{exp2a}
\end{subfigure}
~
~
~
\begin{subfigure}[tbh!]{0.4\columnwidth}
    \resizebox{1\columnwidth}{!}{
    \begin{tikzpicture}
                \begin{axis}[
                    xlabel={Number of Clients},
                    ylabel={Processing Time (S)},
                    symbolic x coords = {2,5,10,15,20},
                    xticklabel style={anchor= east,rotate=45 },
                    xtick=data,
                    ymax=15,
                    ymin=0,
                    legend pos=north west,
                    ymajorgrids=true,
                    grid style=dashed,
                    legend style={nodes={scale=1, transform shape}},
                    label style={font=\Large},
                    tick label style={font=\Large}
                ]
                \addplot+[mark size=3pt]
                    coordinates {
                        (2,1.06)
                        (5,2.46)
                        (10,4.02)
                        (15,4.95)
                        (20,7.02)
                    };
                \addplot+[mark size=3pt]
                    coordinates {
                        (2,2.06)
                        (5,3.46)
                        (10,5.02)
                        (15,6.02)
                        (20,8.92)

                    };
                \legend{Original FL, Fl Aggregation Policy}
                \end{axis}
            \end{tikzpicture}
        }
    \caption{LeNet - CIFAR-10}
    \label{exp2b}
\end{subfigure}
~
~
~
\begin{subfigure}[tbh!]{0.4\columnwidth}
    \resizebox{1\columnwidth}{!}
    {
    \begin{tikzpicture}
                \begin{axis}[
                    xlabel={Number of Clients},
                    ylabel={Processing Time (S)},
                    symbolic x coords = {2,5,10,15,20},
                    xticklabel style={anchor= east,rotate=45 },
                    xtick=data,
                    ymax=15,
                    ymin=0,
                    legend pos=north west,
                    ymajorgrids=true,
                    grid style=dashed,
                    legend style={nodes={scale=1, transform shape}},
                    label style={font=\Large},
                    tick label style={font=\Large}
                ]
                \addplot+[mark size=3pt]
                    coordinates {
                        (2,1.06)
                        (5,1.30)
                        (10,4.22)
                        (15,8.05)
                        (20,10.9)

                    };
                \addplot+[mark size=3pt]
                    coordinates {
                        (2,3.06)
                        (5,3.32)
                        (10,6.22)
                        (15,10.05)
                        (20,12.9)
                    };
                \legend{Original FL,Fl Aggregation Policy}
                \end{axis}
            \end{tikzpicture}
    }
    \caption{AlexNet - MNIST}
    \label{exp2c}
\end{subfigure}
~
~
~
\begin{subfigure}[tbh!]{0.4\columnwidth}
    \resizebox{1\columnwidth}{!}
    {
    \begin{tikzpicture}
                \begin{axis}[
                    xlabel={Number of Clients},
                    ylabel={Processing Time (S)},
                    symbolic x coords = {2,5,10,15,20},
                    xticklabel style={anchor= east,rotate=45 },
                    xtick=data,
                    ymax=15,
                    ymin=0,
                    legend pos=north west,
                    ymajorgrids=true,
                    grid style=dashed,
                    legend style={nodes={scale=1, transform shape}},
                    label style={font=\Large},
                    tick label style={font=\Large}
                ]
                \addplot+[mark size=3pt]
                    coordinates {
                        (2,1.40)
                        (5,2.06)
                        (10,5.22)
                        (15,9.05)
                        (20,11.9)
                    };
                \addplot+[mark size=3pt]
                    coordinates {
                        (2,3.40)
                        (5,4.06)
                        (10,7.22)
                        (15,11.05)
                        (20,13.9)
                    };
                \legend{Original FL, Fl Aggregation Policy}
                \end{axis}
            \end{tikzpicture}
    }
    \caption{AlexNet - CIFAR 10}
    \label{exp2d}
\end{subfigure}
\caption{Processing time of model aggregation process with and without aggregation policy using various machine learning models and datasets.}
\label{exp2}
\end{figure*}

In Fig. \ref{exp2}, we evaluate the performance of our proposed framework for the global model aggregation process. Comparing the original federated learning setup with the policy-based aggregation approach, the experiment shows the global model aggregation time cost difference between the two. We run one round using default federated learning with clients' nodes ranging from two to twenty, and another round of federated learning by enabling the aggregation policy. The results demonstrate the global model aggregation time cost required by LeNet and AlexNet using MNIST and CIFAR-10 datasets.

In Figs. \ref{exp2a} and \ref{exp2b}, the results from a LeNet model performing model aggregation with the default FL setup and policy-based model aggregation using the MNIST and CIFAR-10 datasets are shown. The time cost increases from two to twenty clients for both datasets, with the default model having a maximum time cost of 7 seconds, and our method requiring a maximum of 9 seconds. The average time cost increase is between 1 and 2 seconds compared to the default FL setup. Figs. \ref{exp2c} and \ref{exp2d} show the comparison using an AlexNet model with MNIST and CIFAR-10 datasets. Compared to LeNet, the overall time of AlexNet is higher due to an additional layer, with the maximum additional time cost using our method being 2 seconds.

The results of our experiment demonstrate that the time cost of the original FL setup and policy-based aggregation increases linearly. Moreover, the application of a machine learning model using CIFAR-10 is more time-consuming due to its RGB color. Our proposed aggregation method has a slightly higher time cost than the original FL architecture; however, this slight cost is justified by its secure management system, which leverages policy aggregation smart contracts to record and verify during the global model aggregation process.

\begin{figure}
    \begin{tikzpicture}
      \begin{axis}[
        ybar stacked,,
        ylabel=Deployment Cost (Gas),
        ymin=0,
        ymax=15000000,
        xtick=data,
        xticklabels={Our Method,Saini et al.\cite{9235494},Ouyang et al.\cite{9234516}},
        width=0.45\textwidth,
        height=7cm,
        bar width=1cm,
        enlarge x limits = 0.15,
        ]
        \addplot[fill=blue!50] coordinates {
          (1, 3537625)
          (2, 5783731)
          (3, 10424901)
        };
      \end{axis}
    \end{tikzpicture}
    \caption{Comparison of total gas used during the deployment process}
    \label{exp3}
\end{figure}
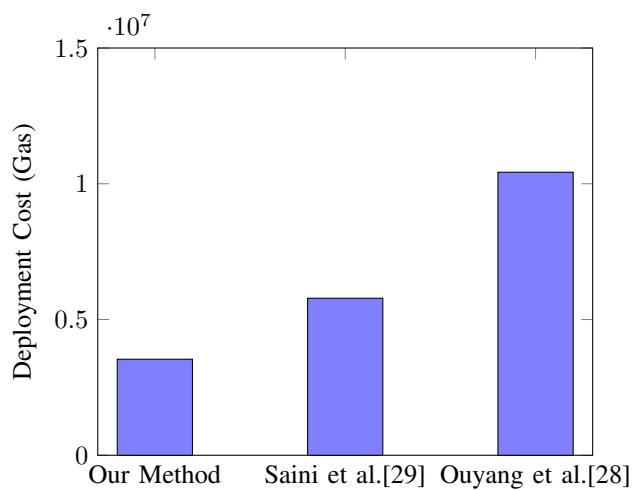

The Ethereum blockchain platform denotes the amount of work done in the form of a unit called gas. In Fig. \ref{exp3}, we compare the total deployment cost (gas) of the smart contract in our proposed framework with Saini et al.\cite{9235494} and Ouyang et al. \cite{9234516}. In this experiment, we deployed the smart contract on the Ethereum blockchain emulator (Ganache) without deploying it in the real Ethereum network. Our proposed method requires 3537625 gas for two smart contracts, while Saini et al. \cite{9235494} requires 5783731 for three smart contracts, and Ouyang et al. \cite{9234516} requires 10424901 gas for three smart contracts. Hence, our proposed method has the lowest gas cost compared to the other approaches. The more information stored on the smart contract, the higher the gas cost during the deployment. For example, Ouyang et al. \cite{9234516} store the models on the smart contract, which results in a high deployment cost. In contrast, we only store the hash of the model on the smart contract so the other party can verify the model based on the hash value stored on the smart contract. In a real-world scenario, these values can be further reduced by using low-cost consensus mechanisms, such as PoS, DPoS, or PBFT.

\subsection{Discussion}
In this section, we summarize the performance of our proposed framework. As discussed in Section \ref{sec:result}, we conducted a series of experiments to assess the effectiveness of our proposed method. Based on the results, the following conclusions can be drawn.

\begin{itemize}
    \item \textbf{Transparency in Management}:
    In this framework, we leverage smart contracts as the underlying technology to develop the training and aggregation policy. The correct utilization of smart contracts for policy control can achieve transparency in the management because all functions executed in the smart contract are reflected on the events log of the smart contract and the Ethereum blockchain network. Therefore, federated learning participants and aggregation nodes can not interrupt the training and aggregation process.
    
    \item \textbf{Local Model Management and Security}:
    In our proposed framework, each participant is subject to the same contract-based training policy, providing a secure management system and ensuring the integrity of the local model from all participants. This policy helps protect against model poisoning attacks, as each $PLM_n$ records the local model hash on the smart contract. Suppose the attacker attempts to tamper with or poison the local model prior to sending it to the cloud for aggregation. In that case, the local model hash will not match the one recorded on the training policy contract and will therefore be rejected. Additionally, the proposed method adds only two minutes to complete the local training with the local training policy, making it a reasonable trade-off for the added security.
    
    \item \textbf{Policy-based Aggregation}:
    In a typical federated learning setup, the aggregation server collects the local training models from each participant and performs the aggregation without verifying the accuracy of the individual local models. In our proposed method, we leverage smart contracts as the base of the aggregation policy to enhance the regular federated learning process. The smart contract will validate the integrity of the sent local model before the aggregation process and records the global model after the aggregation process. The FL participants then use this information when receiving the latest global model. The results indicate that our proposed method only adds a maximum of 2 seconds for executing the policy-based aggregation. Thus, our method provides an effective way to secure the federated learning process while maximizing efficiency.
    
    \item \textbf{Resilience of the Global Model}:
    Blockchain is a compelling and revolutionary decentralized technology that provides data integrity and security by leveraging a secure and resilient network resistant to malicious activities from untrusted parties. Decentralization makes it virtually impossible for attackers to compromise the network, as it would require tampering with every node on the network. In this proposed framework, blockchain technology is utilized to securely store the global model, which has been aggregated from multiple sources. Furthermore, digital signatures and hashes are employed to ensure the integrity of the global model so that attackers cannot modify or corrupt the model, as it would alter the hash value and, consequently, cause the signature verification to fail. Therefore, blockchain technology provides a secure, reliable, and immutable platform for storing the global model, thus ensuring data integrity and enabling distributed decisions.

\end{itemize}

\section{Conclusion}\label{sec:con}

This paper proposes smart policy control for the secure management of federated learning systems. The primary purpose of this work is to ensure that the local models of all participants have the same standard, and each client can verify the integrity of the global model before the next training round. In this framework, we develop a smart policy for local training and aggregation based on a smart contract. The smart contract-based local training policy records the core information when FL participants perform the local training process. The policy control management system then verifies each of the sent local models according to the given training policy contract. Upon successful verification, the local model is sent to the blockchain for aggregation. The aggregation policy records essential information during the aggregation process. Once completed, the global model is securely stored on the blockchain to be subsequently distributed to the federated learning participants in accordance with the policy. After analyzing the experimental results, our proposed method maintains the accuracy of machine learning and provides auditability and verifiability throughout the training and aggregation procedure, albeit with a slight increase in time consumption. Compared to existing work, our methodology has the lowest gas consumption during deployment. Building on this, we plan to develop more efficient blockchain storage and leverage a secure aggregation protocol for additional security measures. Furthermore, we intend to extend our work in the future to support a heterogeneous model, leading to more efficient and secure machine learning.

\section*{Acknowledgment}
This work is supported by the Australian Research Council Discovery Project (DP210102761).

\bibliographystyle{IEEEtran}
\bibliography{References}

\begin{thebibliography}{10}
\providecommand{\url}[1]{#1}
\csname url@samestyle\endcsname
\providecommand{\newblock}{\relax}
\providecommand{\bibinfo}[2]{#2}
\providecommand{\BIBentrySTDinterwordspacing}{\spaceskip=0pt\relax}
\providecommand{\BIBentryALTinterwordstretchfactor}{4}
\providecommand{\BIBentryALTinterwordspacing}{\spaceskip=\fontdimen2\font plus
\BIBentryALTinterwordstretchfactor\fontdimen3\font minus
  \fontdimen4\font\relax}
\providecommand{\BIBforeignlanguage}[2]{{%
\expandafter\ifx\csname l@#1\endcsname\relax
\typeout{** WARNING: IEEEtran.bst: No hyphenation pattern has been}%
\typeout{** loaded for the language `#1'. Using the pattern for}%
\typeout{** the default language instead.}%
\else
\language=\csname l@#1\endcsname
\fi
#2}}
\providecommand{\BIBdecl}{\relax}
\BIBdecl

\bibitem{9903398}
N.~Bugshan, I.~Khalil, M.~S. Rahman, M.~Atiquzzaman, X.~Yi, and S.~Badsha,
  ``Towards trustworthy and privacy-preserving federated deep learning service
  framework for industrial internet-of-things,'' \emph{IEEE Transactions on
  Industrial Informatics}, pp. 1--12, 2022.

\bibitem{9766403}
A.~Hammoud, H.~Otrok, A.~Mourad, and Z.~Dziong, ``On demand fog federations for
  horizontal federated learning in iov,'' \emph{IEEE Transactions on Network
  and Service Management}, vol.~19, no.~3, pp. 3062--3075, 2022.

\bibitem{9997105}
L.~Witt, M.~Heyer, K.~Toyoda, W.~Samek, and D.~Li, ``Decentral and incentivized
  federated learning frameworks: A systematic literature review,'' \emph{IEEE
  Internet of Things Journal}, vol.~10, no.~4, pp. 3642--3663, 2023.

\bibitem{bonawitz2019towards}
K.~Bonawitz, H.~Eichner, W.~Grieskamp, D.~Huba, A.~Ingerman, V.~Ivanov,
  C.~Kiddon, J.~Kone{\v{c}}n{\`y}, S.~Mazzocchi, B.~McMahan \emph{et~al.},
  ``Towards federated learning at scale: System design,'' \emph{Proceedings of
  Machine Learning and Systems}, vol.~1, pp. 374--388, 2019.

\bibitem{9352033}
O.~A. Wahab, A.~Mourad, H.~Otrok, and T.~Taleb, ``Federated machine learning:
  Survey, multi-level classification, desirable criteria and future directions
  in communication and networking systems,'' \emph{IEEE Communications Surveys
  \& Tutorials}, vol.~23, no.~2, pp. 1342--1397, 2021.

\bibitem{yu2021research}
Z.~Yu, S.~U. Amin, M.~Alhussein, and Z.~Lv, ``Research on disease prediction
  based on improved deepfm and iomt,'' \emph{IEEE Access}, vol.~9, pp.
  39\,043--39\,054, 2021.

\bibitem{9403374}
D.~C. Nguyen, M.~Ding, Q.-V. Pham, P.~N. Pathirana, L.~B. Le, A.~Seneviratne,
  J.~Li, D.~Niyato, and H.~V. Poor, ``Federated learning meets blockchain in
  edge computing: Opportunities and challenges,'' \emph{IEEE Internet of Things
  Journal}, vol.~8, no.~16, pp. 12\,806--12\,825, 2021.

\bibitem{zhang2020batchcrypt}
C.~Zhang, S.~Li, J.~Xia, W.~Wang, F.~Yan, and Y.~Liu, ``Batchcrypt: Efficient
  homomorphic encryption for cross-silo federated learning,'' in
  \emph{Proceedings of the 2020 USENIX Annual Technical Conference (USENIX ATC
  2020)}, 2020.

\bibitem{9155414}
X.~Zhang, F.~Li, Z.~Zhang, Q.~Li, C.~Wang, and J.~Wu, ``Enabling execution
  assurance of federated learning at untrusted participants,'' in \emph{IEEE
  INFOCOM 2020 - IEEE Conference on Computer Communications}, 2020, pp.
  1877--1886.

\bibitem{monrat2019survey}
A.~A. Monrat, O.~Schel{\'e}n, and K.~Andersson, ``A survey of blockchain from
  the perspectives of applications, challenges, and opportunities,'' \emph{IEEE
  Access}, vol.~7, pp. 117\,134--117\,151, 2019.

\bibitem{zou2019smart}
W.~Zou, D.~Lo, P.~S. Kochhar, X.-B.~D. Le, X.~Xia, Y.~Feng, Z.~Chen, and B.~Xu,
  ``Smart contract development: Challenges and opportunities,'' \emph{IEEE
  Transactions on Software Engineering}, vol.~47, no.~10, pp. 2084--2106, 2019.

\bibitem{9737334}
J.~Qi, F.~Lin, Z.~Chen, C.~Tang, R.~Jia, and M.~Li, ``High-quality model
  aggregation for blockchain-based federated learning via reputation-motivated
  task participation,'' \emph{IEEE Internet of Things Journal}, vol.~9, no.~19,
  pp. 18\,378--18\,391, 2022.

\bibitem{ohrimenko2016oblivious}
O.~Ohrimenko, F.~Schuster, C.~Fournet, A.~Mehta, S.~Nowozin, K.~Vaswani, and
  M.~Costa, ``{Oblivious Multi-party Machine Learning on Trusted Processors},''
  in \emph{25th $\{$USENIX$\}$ Security Symposium ($\{$USENIX$\}$ Security
  16)}, 2016, pp. 619--636.

\bibitem{9708971}
E.~Kuznetsov, Y.~Chen, and M.~Zhao, ``Securefl: Privacy preserving federated
  learning with sgx and trustzone,'' in \emph{2021 IEEE/ACM Symposium on Edge
  Computing (SEC)}, 2021, pp. 55--67.

\bibitem{juvekar2018gazelle}
C.~Juvekar, V.~Vaikuntanathan, and A.~Chandrakasan, ``{$\{$GAZELLE$\}$: A Low
  Latency Framework for Secure Neural Network Inference},'' in \emph{27th
  $\{$USENIX$\}$ Security Symposium ($\{$USENIX$\}$ Security 18)}, 2018, pp.
  1651--1669.

\bibitem{nakamoto2008bitcoin}
S.~Nakamoto, ``Bitcoin: A peer-to-peer electronic cash system,''
  \emph{Decentralized business review}, p. 21260, 2008.

\bibitem{9617624}
U.~Majeed, L.~U. Khan, A.~Yousafzai, Z.~Han, B.~J. Park, and C.~S. Hong,
  ``St-bfl: A structured transparency empowered cross-silo federated learning
  on the blockchain framework,'' \emph{IEEE Access}, vol.~9, pp.
  155\,634--155\,650, 2021.

\bibitem{9831779}
Y.~Chen, Y.~Zhang, S.~Wang, F.~Wang, Y.~Li, Y.~Jiang, L.~Chen, and B.~Guo,
  ``Dim-ds: Dynamic incentive model for data sharing in federated learning
  based on smart contracts and evolutionary game theory,'' \emph{IEEE Internet
  of Things Journal}, vol.~9, no.~23, pp. 24\,572--24\,584, 2022.

\bibitem{ali2021security}
A.~Ali, H.~A. Rahim, M.~F. Pasha, R.~Dowsley, M.~Masud, J.~Ali, and M.~Baz,
  ``{Security, Privacy, and Reliability in Digital Healthcare Systems Using
  Blockchain},'' \emph{Electronics}, vol.~10, no.~16, p. 2034, 2021.

\bibitem{9686048}
S.~K. Lo, Y.~Liu, Q.~Lu, C.~Wang, X.~Xu, H.-Y. Paik, and L.~Zhu, ``Toward
  trustworthy ai: Blockchain-based architecture design for accountability and
  fairness of federated learning systems,'' \emph{IEEE Internet of Things
  Journal}, vol.~10, no.~4, pp. 3276--3284, 2023.

\bibitem{9930843}
I.~A. Ridhawi, M.~Aloqaily, A.~Abbas, and F.~Karray, ``An intelligent
  blockchain-assisted cooperative framework for industry 4.0 service
  management,'' \emph{IEEE Transactions on Network and Service Management},
  vol.~19, no.~4, pp. 3858--3871, 2022.

\bibitem{9420107}
R.~Kumar, A.~A. Khan, J.~Kumar, Zakria, N.~A. Golilarz, S.~Zhang, Y.~Ting,
  C.~Zheng, and W.~Wang, ``Blockchain-federated-learning and deep learning
  models for covid-19 detection using ct imaging,'' \emph{IEEE Sensors
  Journal}, vol.~21, no.~14, pp. 16\,301--16\,314, 2021.

\bibitem{9822973}
X.~Wang, M.~Peng, H.~Lin, Y.~Wu, and X.~Fan, ``A privacy-enhanced multiarea
  task allocation strategy for healthcare 4.0,'' \emph{IEEE Transactions on
  Industrial Informatics}, vol.~19, no.~3, pp. 2740--2748, 2023.

\bibitem{9170559}
Y.~Zhao, J.~Zhao, L.~Jiang, R.~Tan, D.~Niyato, Z.~Li, L.~Lyu, and Y.~Liu,
  ``{Privacy-Preserving Blockchain-Based Federated Learning for IoT Devices},''
  \emph{IEEE Internet of Things Journal}, vol.~8, no.~3, pp. 1817--1829, 2021.

\bibitem{9285303}
X.~Guo, Z.~Liu, J.~Li, J.~Gao, B.~Hou, C.~Dong, and T.~Baker, ``Verifl:
  Communication-efficient and fast verifiable aggregation for federated
  learning,'' \emph{IEEE Transactions on Information Forensics and Security},
  vol.~16, pp. 1736--1751, 2021.

\bibitem{9321132}
Z.~Peng, J.~Xu, X.~Chu, S.~Gao, Y.~Yao, R.~Gu, and Y.~Tang, ``Vfchain: Enabling
  verifiable and auditable federated learning via blockchain systems,''
  \emph{IEEE Transactions on Network Science and Engineering}, vol.~9, no.~1,
  pp. 173--186, 2022.

\bibitem{10035023}
L.~Ouyang, F.-Y. Wang, Y.~Tian, X.~Jia, H.~Qi, and G.~Wang, ``Artificial
  identification: A novel privacy framework for federated learning based on
  blockchain,'' \emph{IEEE Transactions on Computational Social Systems}, pp.
  1--10, 2023.

\bibitem{9234516}
L.~Ouyang, Y.~Yuan, and F.-Y. Wang, ``Learning markets: An ai collaboration
  framework based on blockchain and smart contracts,'' \emph{IEEE Internet of
  Things Journal}, vol.~9, no.~16, pp. 14\,273--14\,286, 2022.

\bibitem{9235494}
A.~Saini, Q.~Zhu, N.~Singh, Y.~Xiang, L.~Gao, and Y.~Zhang, ``A
  smart-contract-based access control framework for cloud smart healthcare
  system,'' \emph{IEEE Internet of Things Journal}, vol.~8, no.~7, pp.
  5914--5925, 2021.

\bibitem{lecun1998gradient}
Y.~LeCun, L.~Bottou, Y.~Bengio, and P.~Haffner, ``{Gradient-based Learning
  Applied to Document Recognition},'' \emph{Proceedings of the IEEE}, vol.~86,
  no.~11, pp. 2278--2324, 1998.

\bibitem{krizhevsky2017imagenet}
A.~Krizhevsky, I.~Sutskever, and G.~E. Hinton, ``{ImageNet classification with
  deep convolutional neural networks},'' \emph{Communications of the ACM},
  vol.~60, no.~6, pp. 84--90, 2017.

\bibitem{mcmahan2017communication}
B.~McMahan, E.~Moore, D.~Ramage, S.~Hampson, and B.~A. y~Arcas,
  ``{Communication-efficient Learning of Deep Networks from Decentralized
  Data},'' in \emph{Artificial intelligence and statistics}.\hskip 1em plus
  0.5em minus 0.4em\relax PMLR, 2017, pp. 1273--1282.

\bibitem{paszke2019pytorch}
A.~Paszke, S.~Gross, F.~Massa, A.~Lerer, J.~Bradbury, G.~Chanan, T.~Killeen,
  Z.~Lin, N.~Gimelshein, L.~Antiga \emph{et~al.}, ``{Pytorch: An Imperative
  Style, High-performance Deep Learning Library},'' \emph{Advances in neural
  information processing systems}, vol.~32, pp. 8026--8037, 2019.

\bibitem{lee2019testing}
W.-M. Lee, ``Testing smart contracts using ganache,'' in \emph{Beginning
  Ethereum Smart Contracts Programming}.\hskip 1em plus 0.5em minus 0.4em\relax
  Springer, 2019, pp. 147--167.

\bibitem{krizhevsky2010convolutional}
A.~Krizhevsky and G.~Hinton, ``{Convolutional Deep Belief Networks on
  CIFAR-10},'' \emph{Unpublished manuscript}, vol.~40, no.~7, pp. 1--9, 2010.

\bibitem{deng2012mnist}
L.~Deng, ``{The MNIST Database of Handwritten Digit Images for Machine Learning
  Research},'' \emph{IEEE Signal Processing Magazine}, vol.~29, no.~6, pp.
  141--142, 2012.

\end{thebibliography}

\end{document}